\newcommand{\beq}{\begin{equation}}
\newcommand{\eeq}{\end{equation}}
\newcommand{\beqa}{\begin{eqnarray}}
\newcommand{\eeqa}{\end{eqnarray}}
\documentstyle[aps,prl,twocolumn,amsfonts,epsf]{revtex}
\begin{document}
\def\dfrac#1#2{{\displaystyle{#1\over#2}}}

\twocolumn[\hsize\textwidth\columnwidth\hsize\csname @twocolumnfalse\endcsname

\title{Superconductivity Fluctuations in a One-dimensional Two Band
Electron-Phonon Model with Strong Repulsive  Interactions.}

\author{S.I.Matveenko$^{+}$*, A.R.Bishop*  and A.Balatsky*$^{+}$}
\address{$^{+}$ Landau Institute for Theoretical Physics, Moscow, Kosygina
str., 2, Russia, 117940\\
 **Theoretical Division, Los Alamos National Laboratory, Los Alamos,
New Mexico 87545, USA}
\date{April  25, 1995}

\maketitle

\begin{abstract}
We study a one-dimensional, two band model with short range
electron-electron repulsions (onsite $U$ and nearest-neighbour $V$
terms) and electron-phonon coupling. We show that there is a region of
$U$,$V$ and band filling in which singlet superconductivity
fluctuations are dominant. This region is absent without
electron-phonon interactions and includes large values of $U$,$V$.


\end{abstract}


]



The physics of low dimensional strongly correlated fermion systems
with repulsive interactions is a topic of active interest, largely due
to the absence of a clear understanding of the origin of high-$T_{c}$
superconductivity in cuprate oxides and the role of phonons in these
correlated systems.  In the framework of a simple one-dimensional
(1-d) Cu-O chain model \cite{cuo}, we investigate the effects of both
short range electron-electron (e-e) repulsive interactions (onsite $U$
and nearest neighbour Cu-O repulsion $V$) and electron-phonon (e-p)
coupling on the ground state of the system.  We show that
superconducting (SC) correlations are absent in the model if we take
into account e-e interactions only. The inclusion of e-p interactions
leads to the appearance of a region of ($U$, $V$,$\rho$) ($\rho$, the
band - filling) in which superconducting fluctuations are dominant. On
the other hand the ground state of the system in the absence of e-e
repulsion is a charge-density -wave (or spin-density-wave) state
without a divergent SC response.  Thus, the origin of the region with
dominant SC response is the {\it combined} effect of e-e and e-p
interactions for this model.

We use a renormalization group (RG) two cut-off approach, developed in
earlier works \cite{rg,lut}. With some assumptions on the model
parameters our analysis is valid in the large $U,V$ limit.  The
possibility of superconducting fluctuations in quasi-1-d systems with
strong repulsive e-e interactions and e-p coupling was first raised in
work of Zimanyi et.al. \cite{lut}, where results are obtained for a
massive Thirring model. The two-band model without e-p interactions
was considered in \cite{cuo}, where numerical results are presented,
pointing out the existence of SC fluctuations in a strong coupling
limit.


We consider a chain consisting of two types of atoms: Cu on odd sites
with d-orbitals and O on even lattice sites with p-orbitals. The
Hamiltonian of the system is
\beq
H = H_{0} + H_{ee} + H_{ep}
\label{e1}
\eeq

\beq
H_{0} = -t\sum_{<i,j>} c_{p,i}^{+} c_{d,j} + h.c. + \sum_{i} \Delta
(c_{p,i}^{+} c_{p,i} - c_{d,i}^{+} c_{d,i})
\label{e2}
\eeq
\beq
H_{ee} = \sum_{\alpha =d,p}\sum_{i} U_{\alpha} c^{+}_{\alpha,i,\uparrow}
c_{\alpha,i,\uparrow} c_{\alpha,i, \downarrow}^{+}  c_{\alpha,i,\downarrow} +
V \sum_{<i,j>} c_{d,i}^{+}  c_{d,i}  c_{p,j}^{+}  c_{p,j},
\label{e3}
\eeq
where $t$ is the hopping integral, $<i,j>$ are nearest-neighbour
sites, $\Delta = (E_{p} - E_{d})/2$, $E_{p}$ and $E_{d}$ are site
energies, $U_{d}$, $U_{p}$ are Hubbard onsite repulsive energies, and
$V$ is the repulsion amplitude between nearest neighbour sites. Direct
antiferromagnetic coupling betveen Cu sites is omitted. Also

\beq
H_{ep} = H_{ep,1} + H_{ep,2},
\label{e4}
\eeq
and we consider two models of electron-phonon coupling: the molecular
crystal (MC) model with the Hamiltonian $H_{ep,1}$ in which optical
phonons couple to the electron site energy; and the
Su-Schrieffer-Heeger (SSH) model with the Hamiltonian $H_{ep,2} $ in
which the lattice distortions modulate the electron-hopping matrix
element $t$. The Hamiltonian $H_{ep,1}$ consists of two parts
$H_{ep,1} = H_{ep,d} + H_{ep,p}$, where the each part has the form

\beqa
H_{ep} = \sum \frac{P_{i}^{2}}{2M} + \frac{1}{2}\kappa q_{i}^{2} + \lambda
q_{i}\rho_{i} = \nonumber \\
\sum \omega_{0} (d_{k}^{+} d_{k} + \frac{1}{2}) + \frac{g}{N^{1/2}} (d_{k} +
d_{k}^{+}) \rho_{k},
\label{e5}
\eeqa
with $\omega_{0} = (\kappa /M)^{1/2}$, $g =
\lambda/(2M\omega_{0})^{1/2}$, $\rho_{k} =\sum c_{k+q}^{+} c_{q}$.
($M$ ion mass, $\omega_{0}$ optic phonon frequency, $\kappa$
elasticity constant, $\lambda$ e-p coupling constant.) All terms in
(\ref{e5}) have indices $d$ or $p$ and the sum is over odd or even
sites for $H_{ep,d}$, $H_{ep,p}$ respectively.  The Hamiltonian
$H_{ep,2}$ takes into account intermolecular phonon modes

\beqa
H_{ep,2} = \sum \frac{P_{i}^{2}}{2M} + \frac{1}{2}\kappa (q_{i+1} - q_{i})^{2}
-\sum_{<i,j>} \delta t_{i,j} c_{d,i}^{+} c_{p,j} = \nonumber \\
\sum \omega_{k} (f_{k}^{+} f_{k} + \frac{1}{2})
 + \frac{1}{N^{1/2}} \sum  g(k,q)(f_{q} + f_{-q}^{+}) c_{d,k+q}^{+}
c_{p,k},\nonumber \\
\label{e6}
\eeqa
where $\delta t_{i,j} = \lambda (q_{i} - q_{j})$, $\omega_{q} =
2(\kappa /M)^{1/2} \sin (qa/2)$ is the acoustic phonon frequency,
$g(k,q) = i 4\lambda \sin (qa/2) \cos (ka +
qa/2)/(2M\omega_{q})^{1/2}$, and $a$ is the Cu-O lattice constant.

First we consider the noninteracting Hamiltonian  $H_{0}$. Diagonalization
gives
\beq
H_{0} = \sqrt{4t^{2} \cos^{2}ka + \Delta^{2}} (c_{2}^{+}(k)  c_{2} (k) -
c_{1}(k)^{+} c_{1}(k)),
\label{e7}
\eeq
where
\beqa
 c_{d}(k)&  =\cos \theta_{k} c_{1}(k) + \sin \theta_{k} c_{2}(k) \nonumber \\
       c_{p}(k )&  = -\sin \theta_{k}c_{1}(k) + \cos \theta_{k} c_{2}(k)
  \label{e8}
\eeqa
with $\tan (2\theta_{k}) = -2t\cos (ka) /\Delta$, $-\pi /2 <
2\theta_{k} < \pi /2$.  We now have a 2-band electronic structure and
consider the case of an entirely filled lower band. The filling factor
of the upper band is $0 < \rho < 2$ ( empty for $\rho = 0$ and filled
for $\rho = 2$).  With unit cell $2a$, the quasimomentums $k$ and
$k+\pi /a$ are equivalent, and we may consider that states in the
lower band have quasimomenta in the interval $-\pi /2a < k < \pi /2a$
and in the upper band $\pi /2a < \vert k\vert < \pi /a$; then $ k_{F}
a = \pi /2 + \pi \rho /4$. The Fermi velocity is
\beq
v_{F} =  -\frac{2at \sin (2k_{F}a)}{\sqrt{4t^{2}\cos^{2} (k_{F}a ) +
\Delta^{2}}}.
\label{e9}
\eeq

Since we will use an RG approach we take into account only states in
the upper band in the vicinity of $E_{F}$ which are described by
operators $c_{2}$.  Then $H_{0}$ has the form, in $x$-representation,

\beq
H_{0} = v_{F}\Psi_{2,+}^{+} (-i\frac{\partial}{\partial x})\Psi_{2,+} +
v_{F}\Psi_{2,-}^{+} (i\frac{\partial}{\partial x})\Psi_{2,-} ,
\label{e10}
\eeq
where $\Psi_{2,\pm}$ includes momenta near $\pm k_{F}$, respectively.
Below we will omit the index $2$ and also terms in $H$ comprising
$\Psi_{1}$ (taking into account terms with $\Psi_{1}$ can produce a
shift of the chemical potential and some renormalization of the Fermi
velocity.)  Therefore we can substitute in the Hamiltonian terms
\beq
\Psi_{d} \rightarrow \sin \theta_{F} \Psi (x),\;\Psi_{p} \rightarrow \cos
\theta_{F} \Psi (x).
\label{e11}
\eeq
Note that in the cases $t/\Delta \ll 1$ or $\rho \ll 1$ we have
\beq
\sin \theta_{F} \approx \theta_{F} \approx  \frac{t\sin (\pi \rho /4)}{\Delta}.
\label{e12}
\eeq

We first consider e-e interaction effects. For the Cu-O case it is
appropriate to consider $U_{d} \gg U_{p} $.  We first study the case
$U_{p} = 0$.  The effect of small $U_{p}$ is easily taken into account
and will be discussed below. In terms of a "g-ology" model \cite{sol}
the Hamiltonian $H_{ee}$ gives the scattering amplitudes
\beqa
g_{1} = \frac{Ua}{2} \sin^{4} \theta_{F} + 2Va \sin^{2}\theta_{F} \cos (2k_{F}
a) = g_{3} \nonumber \\
g_{2} = \frac{Ua}{2} \sin^{4} \theta_{F} + 2Va \sin^{2}\theta_{F} = g_{4},
\label{e13}
\eeqa
where $g_{1}$ is the back scattering amplitude, and $g_{2}, g_{4}$ are
forward scattering. The "Umklapp" part $g_{3}$ exists only for the
half filled case $\rho = 1$: for the sake of simplicity we will not
consider this case.  Since we use a RG approach below, we consider
$g_{i} /\pi v_{F} \leq 1$, that is $Ua, Va \leq \pi v_{F}$ or $\sin
\theta \ll 1$ for large $U, V$.  We have spin - rotation invariance,
i.e., $g_{\perp} = g_{\parallel}$. Therefore, if it is not essential
we are omitting spin indices.  The effect of the $g_{4}$ term is taken
into account separately: it simply produces a shift in the velocity of
the spin and charge degrees of freedom: $v_{\sigma} = v_{F} (1 +
g_{4})$, $\; v_{\rho} = v_{F}(1 - g_{4})$

The usual RG equations defining the scaling behaviour of the system are
\cite{sol}
\beq
g^{\prime}_{1} = \frac{1}{\pi v_{\sigma}} g_{1}^{2}
\label{e15}
\eeq
\beq
g_{c} \equiv g_{1} - 2g_{2} = const.
\label{e16}
\eeq
For $g_{1} \geq 0$ the excitation spectrum is gapless $g_{1}
\rightarrow g_{1}^{\ast} = 0$, and there is a gap if $g_{1} < 0$.  The
charge excitation spectrum is gapless if $g_{c} \geq 0$ and has a gap
$\Delta_{\rho}$ if $g_{c} < 0$.  The ground state has a most divergent
singlet (triplet) superconductivity (SC) response when $g_{c} \geq 0,
g_{1} < 0 \; (g_{c} \geq 0, g_{1} \geq 0)$. In our case
\beq
g_{c} = -\frac{Ua}{2} \sin^{4} \theta_{F} + 2Va \sin^{2}\theta_{F}
\cos^{2}\theta_{F}(\cos (2k_{F}) - 2) < 0.
\label{e17}
\eeq
Therefore there is no region ($U,V$) with divergent SC fluctuations.
Possible ground states are charge- or spin-density wave, depending on
the sign $g_{1}$. (This sign can vary due to the $\cos k_{F}a$ term).
We see that in order to obtain SC correlations it is necessary to have
large positive $g_{1}^{\ast}$ or negative $g_{2}^{\ast}$ terms.  As we
will see below this condition can be achieved by taking into account
e-p interactions.

Second order perturbation theory in e-p interaction produces a
retarded e-e interaction \cite{rg} for $\omega$ less than a Debye
frequency $\omega < \omega_{D} \sim (\kappa /M)^{1/2}$. (We consider
the case $\omega_{D} < E_{F}$). The effective e-e interaction can be
described in "g-ology" terminology: $ g_{1,ph} = - 2g^{2}(k_{F},
2k_{F})/\omega_{2k_{F}}$,$\; g_{2,ph} = -2g^{2}(k_{F}, 0)
/\omega_{0}$, $\; g_{3,ph} = g_{1,ph} $ (half-filled band only).  In
the case of the MC model (\ref{e5})
\beq
g_{1,ph} =g_{2,ph} = g_{3,ph} = -\frac{\lambda^{2}}{4\kappa},
\label{e18}
\eeq
whereas the SSH model (\ref{e6}) gives
\beq
g_{1,ph} = g_{3,ph} = -4\frac{\lambda^{2}}{\kappa} (\sin^{2} \theta_{F}
\cos^{2} \theta_{F}).
\label{e19}
\eeq
The parameters $\kappa$, $\lambda$ in (\ref{e18}), (\ref{e19}) are, of
course, different, as well as other similar parameters in Hamiltonians
$H_{ep,1}$, $H_{ep,2}$.  Note that all terms are negative, and
$g_{2'ph}$ is due solely to onsite e-p coupling and does not contain
renormalization terms $\sin \theta_{F}$, $\cos \theta_{F}$. In the
case $\theta \ll 1$ the onsite e-p interaction is dominant.  We now
have two types of e-e interactions with cutoffs $E_{F}$ and
$\omega_{D}$. We thus use the RG procedure \cite{rg,lut} for a two
cut-off model. The one-loop scaling equations (\ref{e15}),(\ref{e16})
for the $g_{i}$ are unaffected by the presence of retarded
interactions. The equations for the $g_{i,ph}$, taking into account
cross terms $g_{i} g_{j,ph}$, were derived in \cite{lut}:
\beq
g_{1,ph}^{\prime} = \frac{1}{\pi v_{F}} (\frac{3}{2} g_{1} + \frac{1}{2} g_{c}
+ g_{1,ph} ) g_{1,ph}
\label{e20}
\eeq
\beq
g_{2,ph}^{\prime} = 0.
\label{e21}
\eeq
We will consider the case $g_{3,ph} = 0$. The integration in
(\ref{e20}),(\ref{e21}) is taken from $E_{F}$ to $\omega_{0} \sim
\omega_{D}(\omega_{0})$, where $\omega_{D}(\omega_{0})$ is the
renormalized value of $\omega_{D}$ \cite{lut}.  As a result the
combined action of different scattering processes is described by
\beq
g_{i}^{T} = g_{i}^{\ast} + g_{i,ph}^{\ast}.
\label{e22}
\eeq
The properties of the system at energies small compared to
$\omega_{0}$ are derived from a model with single interactions
$g^{T}_{i}$ and bandwith $\omega_{0}$.

We now examine the solutions of equations (\ref{e15}), (\ref{e16}),
(\ref{e20}), (\ref{e21}). The initial conditions for
(\ref{e15}),(\ref{e16}) are defined by (\ref{e13}).  The initial
conditions for (\ref{e20}),(\ref{e21}) are defined by
(\ref{e18}),(\ref{e19}). We denote $ g_{1,ph}^{(0)} = -\gamma$, $
g_{2,ph}^{(0)} = -\tilde{\gamma}$.  If $g_{1}^{(0)} \geq 0$ (we will
see that this is the situation in the interesting region), from
(\ref{e12}) we obtain that $g_{1}$ scales toward small positive values
$g_{1}^{\ast} \ll g_{1}^{(0)}$.  Note that in the case $\theta \ll 1$
we have $g^{(0)}_{1,ph} \approx g^{(0)}_{2,ph}$, that is
$\tilde{\gamma} \approx \gamma$. From (\ref{e13}) it follows that
$g_{1}^{\ast} - 2g_{2}^{\ast} = g_{1}^{(0)} - 2g_{(2)}^{0}$.  A
positive derivative in (\ref{e20}) implies that $g_{1,ph}$ scales
toward large negative values. We consider the opposite case
$g^{\prime}_{1,ph} < 0$. Then, at least initially, $g_{1,ph}$ will
scale toward a small negative value. Therefore we demand that \beq
\frac{3}{2} g_{1}^{(0)} + \frac{1}{2} g_{c}^{(0)} + g_{1,ph}^{(0)} > 0,
\label{e23}
\eeq
since $g_{1,ph} < 0$. The inequality (\ref{e23}) can not be valid
throughout the scaling process, since $g_{1}$ scales to small values.
Therefore the value $g_{1,ph}^{\ast} $ may not be very small.  We do
not require $ \vert g_{1,ph}^{\ast} \vert \ll \gamma $; for our
purposes it is sufficient that $ g_{1,ph}^{\ast} > \gamma -
2\tilde{\gamma}$, as we show below.  The value $g_{2,ph}$ is not
scaled as follows from (\ref{e21}), i.e., $g_{2,ph}^{\ast} =
-\tilde{\gamma}$. This value does not contain the renormalization
coefficient $\sin \theta_{F}$.  As a result of scaling we have the
state with $g_{i}^{T} = g_{i}^{\ast} + g_{i,ph}^{\ast}$.  The ground
state of the system with the new scaling amplitudes has dominant
divergent SC susceptibility if
\beq
g_{c}^{T} = g_{1}^{T} -2 g_{2}^{T} = g_{1}^{(0)} - 2g_{2}^{(0)} +
g_{1,ph}^{\ast} + 2\tilde{\gamma}> 0.
\label{e24}
\eeq
Since we suppose that $g_{1,ph}^{T} \approx g_{1,ph}^{\ast} < 0$, we have a
state with spin gap $\Delta_{\sigma}$. Therefore the dominant singularity is
the
singlet SC  response with SC correlation function
\beq
R(x) \sim x^{-\frac{1}{K_{\rho}}}, \; K_{\rho} =\sqrt{\frac{1 + g_{c}^{T}/2\pi
v_{\rho} }{1 - g_{c}^{T}/2\pi v_{\rho}}} > 1.
\label{e25}
\eeq
In this case the CDW response can be divergent with correlation
function $\propto x^{-K_{\rho}}$.  The inequalities
(\ref{e23}),(\ref{e24}) define the region in which singlet SC
correlations are dominant.  In terms of $u = \frac{Ua}{2}\sin^{4}
\theta, \; v = Va\sin^{2} \theta \cos^{2}\theta$ we rewrite
(\ref{e23}),(\ref{e24}) as
\beq
\gamma + 2v(1+ 2\cos (\frac{\pi \rho}{2})) < u < 2\gamma^{\ast} - 2v(2 + \cos
(\frac{\pi \rho}{2}),
\label{e26}
\eeq
where $2\gamma^{\ast} = 2\tilde{\gamma} + g_{1,ph}^{\ast}$.  It is
easy to obtain the solution of (\ref{e26}).  This is the region $ABCD$
in Fig.1 delineated by lines: $u=0, v=0, u=\gamma -2v$,
$u=2\tilde{\gamma} + g_{1,ph}^{\ast} - 2v$.
\begin{figure}
\epsfxsize=3in
\centerline{\epsfbox{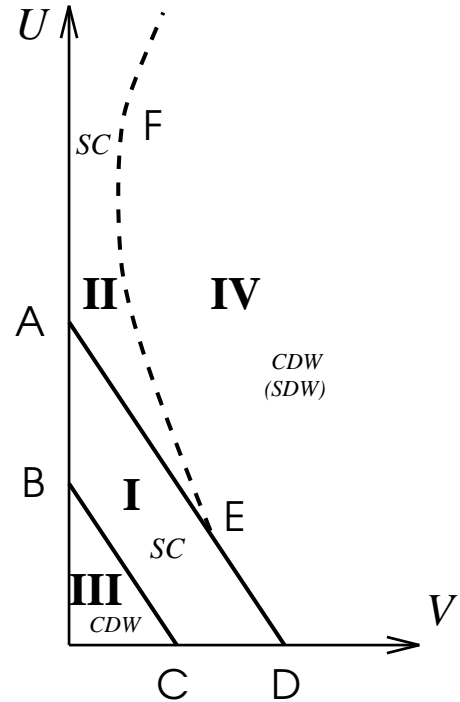}}
\caption{Phase diagram obtained by two cut-off RG scaling. Divergent SC
response regions are I and II.}
\end{figure}
 Recall that in the case $\theta \ll 1$, Eq.(\ref{e12}), the bare
repulsive energies $U \sim u /\theta^{4}, V \sim v/\theta^{2} \gg
\gamma$. Thus our model includes the case of strong electron
repulsion. For any point ($u,v$) in the region ABCD (\ref{e26}) is
valid for
\beq
\rho > \frac{2}{\pi} \cos ^{-1} (max\{\frac{u-\gamma -2v}{4v}, \;
\frac{2\gamma^{\ast} -4v-u}{2v} \},
\label{e27}
\eeq
In the case $t/\Delta \ll 1$ we can obtain the phase diagram in term
of the bare values $U,V$. Then the coordinates of points $A,B,C,D$ are
$A = \{0, (4\gamma^{\ast}) (\Delta /t)^{4} \}$, $B = \{0, 2\gamma
(\Delta /t)^{4} \}$, $C = \{(\gamma /2)(\Delta /t)^{2}, 0\}$, $D =
\{(\gamma^{\ast}/2)(\Delta /t)^{2}, 0\}$.  The SC region is deformed
to include region II due to the $\sin (\pi \rho /4)$ term.  The
equation of the curve EF is
\beq
V = \frac{\Delta^2}{t^2} \frac{((2\gamma^{\ast} - \gamma)k + 4\gamma + 16
\gamma^{\ast})^{2}}{72(k+2)(\gamma + 2 \gamma^{\ast})},\;
U = 2kV\frac{\Delta^2}{t^2}.
\label{e28}
\eeq
In the limit $k \rightarrow \infty$ we have $U \propto V^{2}$, but in
this region $\rho \sim 1/U^{1/4} \rightarrow 0$.  The inequality
(\ref{e26}) with $t/\Delta \ll 1$ becomes
\beqa
(U_1 + 8V_1 )y^2 - 6V_1 y - \gamma > 0, \nonumber \\
(4V_1 - U) y^2 - 6V_1 y + 2\gamma^{\ast} > 0,
\label{e29}
\eeqa
where $V_{1} =V(\Delta /t)^{2}, U_{1} = \frac{U}{2}(\Delta /t)^{4}$,
$y = \sin^{2}(\pi \rho /4)$.  In region I the solution of (\ref{e29})
is $\rho_{0} < \rho < 2$, where $\sin^{2}(\pi \rho_{0} /4) = {y_{0}}$
is the largest root of eqn. (\ref{e29}). In the region II we have
$\rho_{1}(U,V) < \rho < \rho_{2}(U,V)$, where $\rho_{1}$, $\rho_{2}$
are easy obtained from (\ref{e29}).  If $V=0$ the solution is
\beq
\frac{4}{\pi} \sin^{-1} \left(\frac{\gamma}{U_{1}}\right)^{1/4} < \rho <
\frac{4}{\pi} \sin^{-1} \left(\frac{2\gamma^{\ast}}{U_{1}}\right)^{1/4}
\label{e31}
\eeq
for $U_{1} > \gamma$; if $U=0$ , then in the region $\gamma /2 < V_{1}
< \gamma^{\ast}$ the solution is $\sin^{2}(\pi \rho /4) > y_{0}$,
where $y_{0}$ is the largest root of eqn. (\ref{e29}) for $U=0$.

In using the RG approach, we supposed as usual that $g_{i} /\pi v_{F}
< 1$.  For small $t/\Delta$ we have the initial value $v_{F}^{(0)}
\sim (t/\Delta)\sin(\pi \rho /2)$. Recalling that $g_{i} \sim (t\sin
(\pi \rho /4) /\Delta)^{2} $ or $(t\sin (\pi \rho /4) /\Delta)^{4}$,
$g_{2,ph} = const$, we suppose that our results are reasonable if we
are not too close to band edges, that is $\epsilon_{1}< \rho < 2 -
\epsilon_{2}$, and $\rho \neq 1$ ($g_{3} = 0$).  It follows from our
analysis that in region III we have large spin and charge gaps, so
that there is only a charge-density-wave divergent response.  In the
region IV we have $g_{c}^{T} < 0$ , small $g_{1}^{T} < 0$ and thus
divergent charge- and spin-density-wave (if $g_{1}^{\ast} \rightarrow
0$) responses.

In our consideration above we have not taken into account effects of
$U_{p}$ repulsion. This is easily achieved by substituting $u$ into
(\ref{e21}) in the form $u = (U a\sin^{4}\theta_{F} + U_{p}a
\cos^{4}\theta_{F})/2$.  For small values of $U_{p}$, the RG approach
remains valid, and all results in terms of the new $u,v$ are retained.
For $t/\Delta \ll 1$ we have $\cos \theta_{F} \sim 1$, so that we can
not consider the large $U_{p}$ limit in our approach.

In conclusion, using a two cutoff RG approach we have studied a two
band, 1-d tight-binding model with e-e and e-p interactions.  We
included onsite $U$ and nearest neighbour $V$ electron repulsions, as
well as intra- and inter-molecular e-p coupling.  We have shown that
there is no $U,V,\rho, t,\Delta$ parameter region with dominant
divergent SC response in the absence of e-p interactions.  In the
lowest order RG approach we found that such a region {\it does} occur
if we include e-p coupling with optical intra-molecular modes. Only
this form of e-p interaction produces an effective renormalized
$g_{2,ph}$ term.  We have found that the singlet SC region includes
large values of the $U,V$ repulsive interactions if $t\sin(\pi \rho
/4) /\Delta \ll 1$. Note that this behavior is impossible in the
framework of a one-band model, for which $\Delta = 0$.  Then, instead
of (\ref{e26}) we have
\beq
\gamma + 2V(1- 2\cos (\pi \rho)) < U < 2\tilde{\gamma} + g_{1,ph}^{\ast} - 2V(2
- \cos  (\pi \rho),
\label{e33}
\eeq
where $\gamma = -g^{(0)}_{1,ph}$, $\tilde{\gamma} = -g^{(0)}_{2,ph}$,
$0 <\rho <2$.  The solution of (\ref{e33}) is the same region ABCD in
Fig.1, provided $2\tilde{\gamma} + g^{\ast}_{1,ph} > \gamma$. However
the bare values $U,V$ must be very small, of the order of phonon
scattering strengths. Moreover, in this case the requirement
$2\tilde{\gamma} + g^{\ast}_{1,ph} > \gamma$ can be not realized.
(Recall that in the two band case we have $\gamma \sim \tilde{\gamma}$
in the limit $t/d \ll 1$, since effective intersite phonon coupling
terms are small due to the factor $(t/\Delta)^{2}$ ).  Note also that
we have used a RG approach. Therefore we did not consider the strong
coupling limit where a phase separation instability could take place
\cite{cuo}: our results are valid in some vicinity of the SC
fluctuation phase.

We have proposed one possible scenario for the origin of dominant SC
fluctuations in quasi-one-dimensional systems as a result of the
combined effect of repulsive e-e and attractive e-p interactions in a
2-band situation. We suggest that features of this picture will
survive in analogous two-dimensional models of high-$T_{c}$
superconductors \cite{tes}, in particular in 3-band Peierls-Hubbard
models. However the orbital structure of the order parameter in this
case (s-wave vs d-wave) is unclear without detailed calculations.

One of us (S.M) wishes to thank Los Alamos National Laboratory for
support and hospitality. Work at Los Alamos is performed under the
auspices of the U.S.D.o.E.

\end{document}